Engineering tunnel junctions on ballistic semiconductor nanowires

J. Damasco[1], S. T. Gill[1], S. Gazibegovic[2,3], G. Badawy[2], E. P. A. M. Bakkers[2,3], N. Mason[1]

1. Department of Physics, University of Illinois at Urbana−Champaign, Urbana, Illinois 61801, United States
2. Department of Applied Physics, Eindhoven University of Technology, 5600 MB, Eindhoven, The Netherlands
3. QuTech and Kavli Institute of NanoScience, Delft University of Technology, 2600 GA, Delft, The Netherlands

Abstract

Typical measurements of nanowire devices rely on end-to-end measurements to reveal mesoscopic phenomena such as quantized conductance or Coulomb blockade. However, creating nanoscale tunnel junctions allows one to directly measure other properties such as the density of states or electronic energy distribution functions. In this paper, we demonstrate how to realize uniform tunnel junctions on InSb nanowires, where the low invasiveness preserves ballistic transport in the nanowires. The utility of the tunnel junctions is demonstrated *via* measurements using a superconducting tunneling probe, which reveal non-equilibrium properties in the open quantum dot regime of an InSb nanowire. The method for high-quality tunnel junction fabrication on InSb nanowires is applicable to other III-V nanowires and allows for new tools to characterize the local density of states.

Main

InSb nanowires provide a versatile platform for novel electrical devices as a result of high mobility[1], large spin-orbit coupling[2], and the ability to be extended to nanowire networks[3]. Previous measurements of InSb nanowires device have focused on end-to-end transport measurements, which have demonstrated coherent mesoscopic phenomena such as quantized conductance[4], quantum interference[3], and Josephson effect[5]. However, new understanding of these materials can be gained by utilizing tunneling spectroscopy, which can directly probe the electronic density of states (DOS). For example, a long standing goal in the study of ballistic, semiconducting nanowires with strong spin-orbit coupling is to demonstrate helical modes, which can provide the basis for novel spintronic applications[6] and engineering topological superconductivity[7]. While it has been a challenge to detect helical modes using end-to-end transport measurements as a result of sensitivity to chemical potential variations[8], it may be possible to detect these modes using momentum-resolved tunneling spectroscopy[6,9].

Despite the interest in using InSb in novel electrical devices, the basic features such as the DOS and energy dispersion are not known *a priori* and determining these features for individual wires is challenging. For example, InSb nanowires are not well-suited for techniques such as scanning tunneling spectroscopy as a result of technical difficulties arising from the surface sensitivity[10]. More accessible and device-compatible tunneling spectroscopy can be achieved by fabricating planar tunnel junctions, where a metal probe is separated from the nanowire by a thin insulating barrier. Further, the use of a superconducting probe in a planar tunnel junction enables direct measurements of electron distribution functions and interactions[11], along with enhanced spectroscopy of mesoscopic tunneling effects[12] and bound states[13,14].

To perform tunneling spectroscopy experiments where ballistic transport is maintained under the probe, high-quality tunnel barriers in conjunction with careful surface engineering of the nanowire is required to prevent the introduction of disorder. Aluminum is often used for tunnel junctions because the metal develops a well-controlled and self-limiting oxide layer. However, many semiconducting nanowires, including InSb, already have intrinsic surface oxides that are too thick for tunneling spectroscopy and typically contain a high defect density. Therefore the native oxide layer needs to be removed— without introducing disorder into the wire—and replaced with a higher quality barrier material. In this Letter, the systematic development of high-quality $AlO_x$ tunnel barriers on InSb nanowires is described. Through the use of controlled etching, deposition, and oxidation techniques, nanowires having a low-roughness interface for the growth of pinhole-free $AlO_x$ tunnel barriers are realized. The quality of tunnel barrier fabrication is characterized using superconducting tunneling spectroscopy, which reveals a hard superconducting gap consistent with a uniform $AlO_x$ tunnel barrier. Additionally, in device having uniform barriers, ballistic transport features such as Fabry-Perot resonances and quantized conductance steps are observed. Futhermore, the utility of tunnel junctions on InSb nanowires is demonstrated by performing non-equilibrium tunneling spectroscopy of an open quantum dot formed in an InSb nanowire, which reveals the presence of strong electron-electron interactions. The development of low-invasive tunnel junctions on InSb nanowires will allow for new tools to characterize the local density of states, which may find use in momentum resolved tunneling or characterizing the phase-dependence of Andreev bound states in nanowire Josephson junctions.

The devices presented in this paper use InSb nanowires that were grown by metal-organic vapor phase epitaxy[15] and are deterministically transferred to chips having a $SiO_2$ dielectric that was thermally grown on heavily *p*-doped Si. The $Si/SiO_2$ acts as a back gate for the nanowire. The $SiO_2$ substrate is pre-patterned and then scrubbed of polymer debris using reactive ion etching prior to nanowire deposition. All contacts to the nanowires are defined by electron beam lithography (EBL). The devices reported in this paper were patterned in a three-terminal geometry consisting of two highly-transparent normal metal contacts to InSb, separated by ~1 μm, and a central superconducting tunnel probe. These devices enable the measurement of both two-terminal transport and tunneling spectroscopy to monitor the DOS in the electron cavity defined by the electrostatic profile of the probe. The transparent contacts are prepared using methods described elsewhere to realize ballistic InSb nanowire devices[4,16,17]. Following contact development, a probe is patterned using EBL, leaving a region of InSb nanowire coated in its native oxide. Following a brief reactive ion etch to remove any polymer remaining from lithography, the native oxide is removed using sulfur-based etching, which has been shown to leave a smooth InSb surface for further processing[5,16]. Prior to barrier deposition, the nanowire is briefly ion-milled to remove adsorbates following the wet etch step. To create the $AlO_x$ barrier, 0.7 Å of Al was deposited at a rate of 0.1 Å/s followed by oxidation in high-purity $O_2$ at a pressure of 10 mTorr for 1 hour. This deposition/oxidation step is then repeated two more times. Finally, a superconducting metal is deposited *in situ* on the barrier.

Figure 1A shows an SEM image of a completed device, while Fig. 1B shows the tunneling spectra from the superconducting probe to the InSb. Superconducting tunneling spectroscopy provides an excellent litmus test for the quality of the barrier because deviations from a BCS-like DOS can typically be linked to "leaks" and inhomogeneity in the barrier region[18]. As shown in Figure 1B, using our optimized recipe, a BCS-like DOS having a hard gap was observed when biasing the tunnel probe lead with one of the normal leads grounded. This indicates a low-leakage, low-disorder junction. It is also important that the deposition is minimally invasive to the nanowire, and ballistic transport can be maintained in the region under the probe. Figure 1C shows end-to-end conductance for device shown in Figure 1A, where the signature "chess-board" pattern of Fabry-Perot interference as a function of gate voltage and source-drain voltage is evident, indicating quasiballistic conduction across the device[19]. The Fabry-Perot patterns can be analyzed to compare the length of the ballistic region to the periodicity of gate voltage oscillations: the change of the Fermi wave vector over one period is given as $\delta k_F = \pi/L$, which is related to a change of carrier density by $2\delta k_F / \pi = C_{G,L} \Delta V_{G,FP}/e$, where $C_{G,L}$ is the gate capacitance per unit length of the cavity and $\Delta V_{G,FP}$ is the periodicity of the Fabry-Perot oscillations. Hence, the cavity length can be estimated as

$$L = \frac{2e}{\Delta V_{G,FP} C_{G,L}} \quad . \tag{1}$$

The gate capacitance in the region of the probe is negligible compared to the capacitance of the quantum point contacts to the gate, which is on the order of 5 aF[1]. Using this approximation, the gate capacitance per unit length of the cavity is $C_{G,L} \approx 10$ aF/μm. Given an average $\Delta V_{G,FP}$ of 55 mV, the cavity length is estimated to be L ~ 600 nm, which is approximately the length from the center of one bare region of InSb to the next, as schematically shown in Figure 1A. Hence, the estimated cavity length agrees with the lithographically-defined pattern, indicating quasiballistic transport along the entire nanowire, even under the superconducting tunnel probe.

The growth parameters that are required to realize a nearly ideal planar barrier, as sketched in Figure 2A, will now be discussed. From our experience[18], an ideal barrier has a room temperature resistance above 100 kΩ, which allows the observation of a BCS-like density of states in the superconducting probe at low temperature. In general, depositing and oxidizing ultrathin films of Al on InSb does not guarantee a continuous barrier. Realistically, pinholes form from incomplete nucleation of $AlO_x$ grains on the nanowire surface, as shown in Figure 2B. This motivated us to use liquid nitrogen to perform a cooled deposition of Al to promote the most continuous layer. However, as shown in Table 1, all barriers fabricated at low temperature and with a sticking layer before depositing the superconductor demonstrated resistances below what is required to observe a BCS-like DOS in tunneling measurements. The cold evaporation of ultrathin barriers likely leaves incomplete coverage on sides from shadowing, as shown in Fig 2C, which leaves a large area for the sticking layer to make a partial ohmic connection to the wire. To suppress any ohmic contact to the wire following barrier deposition, the superconductor was deposited without a sticking layer[5]. In doing so, the deposition of Al or NbTiN produces a high-resistance contact to InSb because of poor wetting of the metal on the semiconductor surface at room temperature[5,16]. After removing the sticking layer from our recipe, both cold and room temperature evaporation of the barrier could provide resistances that produced high-quality tunneling measurements at low temperatures. However, room temperature deposition of the barrier led to a higher yield of functioning tunnel junctions at low temperatures. Table 1 summarizes the conditions for creating tunnel junctions to InSb nanowires.

Figure 3 presents the properties of an optimized, nearly disorder-free device. For the data presented in Figure 3, the tunnel barriers were deposited at room temperature and without a sticking layer. In Figure 3A, the magnitude of the tunnel conductance is independent of gate voltage, and the superconducting gap is seen over the entire scan with high clarity and no features forming below the gap, $\Delta_{Al}$ = 220 μV. Figure 4B shows a tunneling measurement performed at zero gate voltage with the superconducting tunnel probe grounded, demonstrating a BCS-like DOS having an above-gap to subgap conductance ratio greater than 10, consistent with the deposition of a highly uniform tunnel barrier. The gate-independent magnitude of tunneling conductance and absence of bound states observed below the superconducting gap attest to the high quality of the tunnel junction. Only above the gap can spectroscopic features from the nanowire be observed. Hence, these uniform, weakly coupling tunnel junctions allow for the characterization of the DOS in nanowires without significant contributions from probe-nanowire hybridization. Additionally, while this device was unable to fully pinch-off before dielectric breakdown, approximately quantized transport across the three-terminal geometry was observed, as shown in Figure 4C. The observation of roughly quantized conductance further confirms the uniformity of the barrier and low disorder impinged from fabricating the tunnel junction.

Finally, data are shown from an important application of superconducting tunnel probes: non-equilibrium tunneling spectroscopy. As shown in Figure 4A, this measurement involves applying a fixed voltage $U$ across the device to drive the system out of equilibrium. The conductance across the tunnel junction is measured and is given by

$$\frac{\partial I}{\partial V_B}(V_B) = c\int_{-\infty}^{\infty} \frac{\partial n_{probe}(E)}{\partial E} n_{sample}(E - eV_B)\left[f_{sample}(E - eV_B) - f_{probe}(E)\right]dE \quad (2)$$

where $E$ is the energy relative to the Fermi energy, $n_{probe}$ is the superconducting density of states in the probe, $n_{sample}$ is the density of states in the 1D wire, and $f_{probe}$ and $f_{sample}$ are the Fermi distribution functions in the probe and the 1D wire, respectively. This expression[21] is a convolution of the electron energy distribution of the nanowire device with the voltage $U$ with the DOS in the nanowire and the gradient of the DOS of the superconducting probe. Generally, the functional form of $f_{sample}$ depends on how electrons are scattered as they travel down the nanowire. Figure 4B shows that as the non- equilibrium voltage is increased across the device, the gap shifts laterally while the peaks are smoothed out. Similar non-equilibrium behavior has been observed in carbon nanotubes and was associated with electron-electron scattering[12]. Although the device measured in Fig. 4B has non-ideal tunnel barriers, leading to a soft gap, it demonstrates the utilization of non- equilibrium superconducting tunneling spectroscopy on InSb nanowires, and shows the nature of scattering caused by coupling the disorder of the non- ideal barrier into the quantum wire. Indeed, as shown in Figure 4C, simulations of the impact that electron-electron scattering has on non- equilibrium superconducting tunneling spectroscopy possess strong qualitative resemblance to the experimental results. The simulations of the superconducting gap behavior with applied non-equilibrium voltage were performed following the procedure of previous work[12,21,22] and assumed that the nanowire device was thermalized to 250 mK and had a superconducting tunnel junction with $\Delta_{Al}$ = 240 µV.

The methods presented here for fabricating high-quality tunnel junctions on high-mobility nanowires with strong spin-orbit coupling may have immediate applications for future studies of mesoscopic superconductivity. In a modified configuration of the same three-terminal geometry and by replacing the normal leads with supeconducting leads, the phase-dependence of Andreev bound states in InSb nanowire Josephson junction can be studied[14,23]. In addition, our work also lays the groundwork for developing thinner barriers that will enable proximity superconductivity in the wire and prevent deleterious metallization effects[24].

In conclusion, high-quality tunnel junctions to ballistic cavities in InSb nanowire were developed. The performance of various barrier growth parameters and room temperature resistances that correspond to uniform, pinhole-free junctions were identified. Measurements confirmed that minimal disorder is added by the fabrication and that ballistic transport is maintained in the area under the tunnel junction. Superconducting tunneling spectroscopy demonstrated the high quality of the junction, and non-equilbrium tunneling spectroscopy was used to determine the extent of electron scattering. The use of high-quality tunnel junctions on quantum wires following the guidelines in this paper can be used to reveal salient features of helical modes and topological superconductivity in nanowire devices.

**Acknowledgments**
S.T.G, J.D., and N.M. were supported by the Office of Naval Research Grant No. N0014-16-1-2270 and NSF DMR-1710437. S.G, D.C., and E.P.A.M. B. were partly supported by the European Research Council and The Netherlands Organization for Scientific Research. S.T.G acknowledges support from an NSF Graduate Research Fellowship. S.T.G, J.D., and N.M. acknowledge the use of the Materials Research Lab Central Facilities at the University of Illinois for all work.

**Figure 1.** A. SEM image of a typical tunneling spectroscopy device, consisting of an Al probe and two Au leads. The bare contrictions form quantum point contacts (QPCs), which control the coupling of the cavity to the leads. In these experiments, the QPCs are tuned simultaneously using a global backgate. Scale bar in white is 200 nm. B. Superconducting tunneling spectroscopy of the device in Fig 1A at $V_g = 0$ V, showing a BCS-like density of states. The probe is grounded through a current preamp while standard lock-in techniques are used to measure the conductance as a function of voltage bias. The measurement setup is shown schematically in the inset. C. End-to-end conductance of the device in Fig 1A as a function of gate voltage. The color map shows typical Fabry-Perot checkering as a function of gate and bias from interference in a ballistic electron cavity. The line cut from the color map is taken for $V_b = 0$ mV and shows that the oscillations are on the order of $e^2/h$.

**Figure 2.** A Cross-sectional schematic of an ideal, planar tunnel barrier on an InSb nanowire. B Cross-sectional schematic of a leaky, pinholed tunnel barrier. C. Cross-sectional schematic of shadowing effect from cold deposition of an $AlO_x$ barrier.

**Table 1.** List of superconducting materials used for probes, the barrier thickness whether the barrier was deposited cold or a Ti sticking layer was used, and the room temperature resistance range.

**Figure 3.** A. Gate dependence of the tunnel conductance as a function of bias. For entire gate range, the magnitude of the tunnel conductance remains constant, and the superconducting gap is seen over the entire scan with high clarity and no features forming below the gap, $\Delta_{Al}$ = 220 µV. B. Plot of differential conductance as a function of bias for the tunnel junction. A hard gap is observed with an above gap to subgap conductance ratio greater than 10. Electron temperature is estimated as 100-150 mK, resulting in slight thermal rounding of the gap. C. Zero-bias end to-end conductance as a function of gate showing plateaus, all of which are nearly quantized in units of $2e^2/h$. Arrows point to plateau features in the conductance.

**Figure 4.** A. Schematic of non-equilibrium tunneling spectroscopy. A floating voltage, U, is applied across the ends while measuring the tunneling differential conductance of the superconducting probe as a function of bias. B. Tunneling conductance as a function of non-equilibrium voltage. As the non-equilibrium voltage is increased across the device, the gap shifts laterally while the peaks are smoothed out. C. Simulation of the effect a non-equilibrium bias has on 1D wire having strong electron-electron scattering. Simulations show qualitatively similar behavior to the experimental device, implying strong electron-electron scattering in the device.

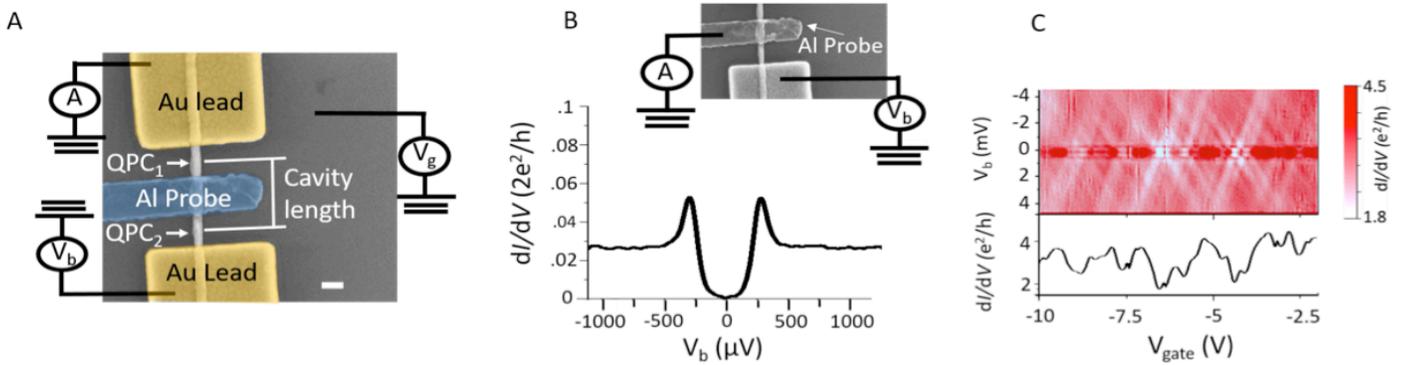

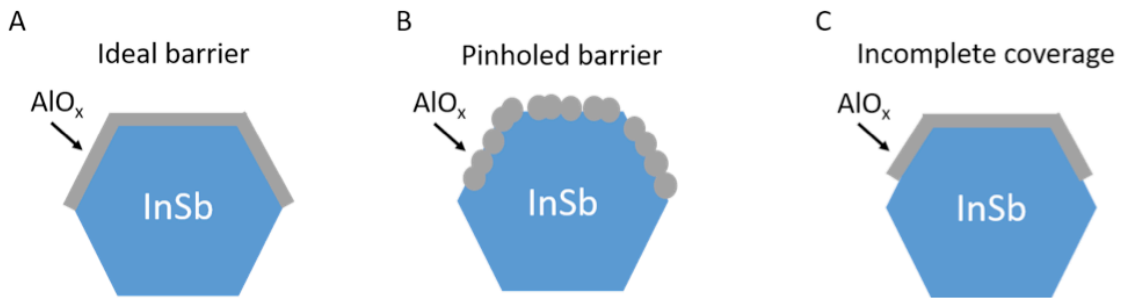

| Probe Material | Cold evaporation of barrier | Ti sticking layer | Barrier thickness | T = 293 K probe resistances |
|---|---|---|---|---|
| Al | Y | Y | 2.1 nm | 20-30 kΩ |
| NbTi/NbTiN | Y | Y | 2.1 nm | 20-40 kΩ |
| Al | Y | Y | 2.7 nm | 60-80 kΩ |
| Al | Y | N | 2.1 nm | 100 kΩ-100+ MΩ |
| Al | N | N | 1.9 nm | 1-100+ MΩ |

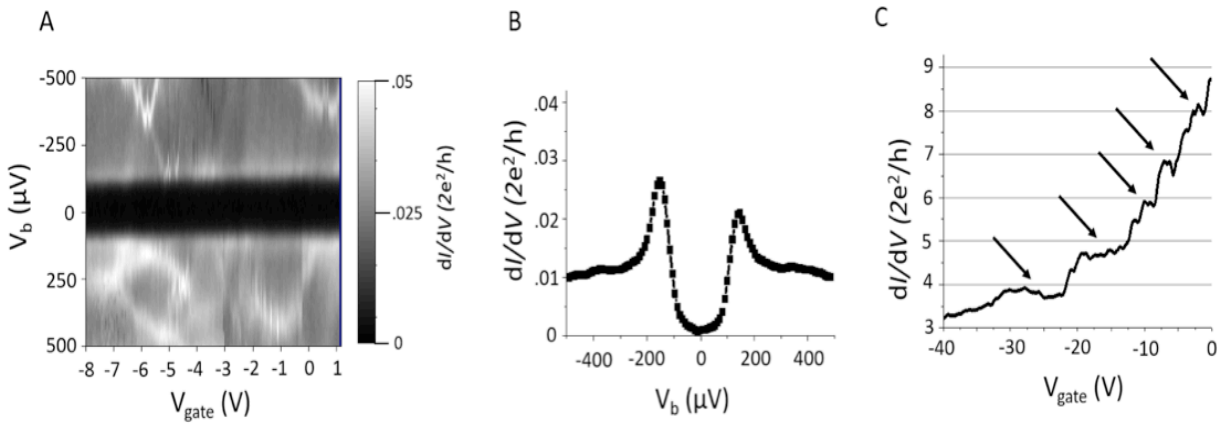

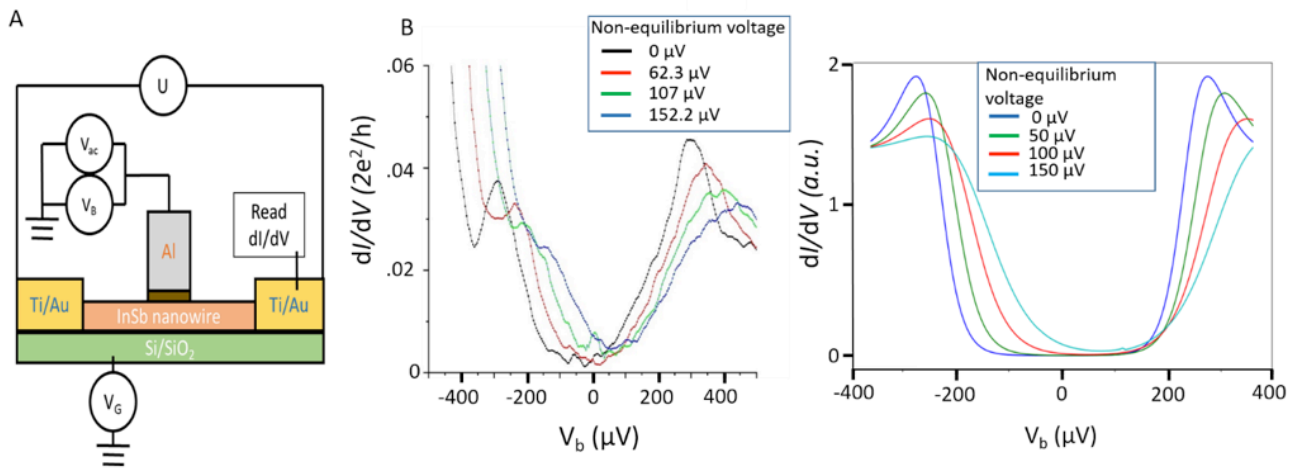